\documentclass[12pt,preprint]{aastex}

\begin{document}

\title{A Major SGR-like Outburst and Rotation Glitch in the No-Longer-So-Anomalous X-ray Pulsar 1E 2259+586}

\author{
V. M. Kaspi,\altaffilmark{1,2,3}
F. P. Gavriil,\altaffilmark{1}
P. M. Woods,\altaffilmark{4}
J. B. Jensen\altaffilmark{5}
M. S. E. Roberts,\altaffilmark{1,2}
D. Chakrabarty\altaffilmark{2}
}

\altaffiltext{1}{Department of Physics, Rutherford Physics Building,
McGill University, 3600 University Street, Montreal, Quebec,
H3A 2T8, Canada}

\altaffiltext{2}{Department of Physics and Center for Space Research,
Massachusetts Institute of Technology, Cambridge, MA 02139}

\altaffiltext{3}{Canada Research Chair}

\altaffiltext{4}{Space Science Research Center, National Space Science
and Technology Center, Huntsville, AL 35805, USA; Universities Space
Research Association}

\altaffiltext{5}{Gemini Observatory, 670 North A'ohoku Place, Hilo, HI 96720}

\begin{abstract}

We report a major outburst from the Anomalous X-ray Pulsar 1E~2259+586,
in which over 80 X-ray bursts were detected in four hours using the
{\it Rossi X-ray Timing Explorer}.  The bursts range in duration from
2~ms to 3~s and have fluences in the 2--10~keV band that range
from $3 \times 10^{-11}$ to $5 \times 10^{-9}$~erg~cm$^{-2}$.  We
simultaneously observed increases of the pulsed and persistent X-ray
emission by over an order of magnitude relative to quiescent levels.
Both decayed significantly during the course of our 14~ks observation.
Correlated spectral hardening was also observed, with the spectrum
softening during the observation.  In addition, we observed a pulse
profile change, in which the amplitudes of the two peaks in the pulse
profile were swapped.  The profile relaxed back to its pre-outburst
morphology after $\sim$6 days.  The pulsar also underwent a sudden
spin-up $(\Delta \nu / \nu = 4 \times 10^{-6}$), followed by a large
(factor of $\sim$2) increase in spin-down rate which persisted for
$>$18~days.  We also observed, using the {\it Gemini}-North
telescope, an infrared enhancement, in which the $K_s$ (2.15~$\mu$m)
flux increased, relative to that measured in a observation made in
2000, by a factor of $\sim$3, three days post-outburst.  The IR counterpart
then faded by a factor of $\sim$2 one week later.  In addition, we
report an upper limit of 50~$\mu$Jy on radio emission at 1.4~GHz two days
post-outburst.  The X-ray properties of this outburst are like those
seen only in Soft Gamma Repeaters.  This conclusively unifies Anomalous
X-ray Pulsars and Soft Gamma Repeaters, as predicted uniquely by the
magnetar model.

\end{abstract}


\section{Introduction}

Radiation from some isolated young neutron stars has been suggested to
be powered by the decay of an ultra-high magnetic field
\citep{td95,td96a}.  Only this ``magnetar'' model has been able to explain the
properties of the rare class of soft gamma repeaters (SGRs),
whose hallmark is repeated X-ray and gamma-ray bursts \citep[e.g.][]{kds+98}.
A different rare class, Anomalous X-ray Pulsars (AXPs), are
characterized by soft X-ray pulsations \citep{ms95}, and have also been
proposed as magnetars \citep{td95}, from the similarity of their pulsations with
those of SGRs in quiescence.  
However, competing models in which AXPs are accretion powered, via a
fall-back disk rather than an unseen binary companion, have been
proposed \citep{chn00}.  Recently, the detection of two weak X-ray
bursts from the direction of AXP 1E~1048.1$-$5937 argued for AXPs being
magnetars \citep{gkw02}.  That AXP had previously been identified as
one that was most likely to burst on the basis of its unstable timing
behavior which was reminiscent of that seen in SGRs, and because of its
SGR-like spectrum \citep{kgc+01}.

1E~2259+586, a 7-s AXP in the supernova remnant CTB~109 \citep{fg81},
in contrast to 1E 1048.1$-$5937, has shown remarkably stable timing
behavior and pulsed X-ray fluxes in the past 5.6~yr
\citep{kcs99,gk02}.  It also has the smallest inferred surface dipolar
magnetic field of all AXPs (and SGRs for which it has been determined),
and has an X-ray spectrum softer than those of the SGRs.  Past
observations have suggested, however, that the pulsar may experience
epochs of activity, including flux, timing, and pulse profile
variations \citep{ikh92,cso+95,bs96}.

Here we report a major SGR-like outburst from the AXP 1E~2259+586, in
which over 80 X-ray bursts were detected along with a variety of
significant changes to the pulsed and persistent emission.

\section{Observations and Results}

The 1E~2259+586 outburst was detected in an observation that was made
as part of a long-term {\it Rossi X-ray Timing Explorer (RXTE)} AXP
monitoring program \citep{kcs99,gk02}.  Unexpectedly, bursts were seen
during a 14.4~ks observation on June 18, 2002 (UT 15:39).  The total on
source exposure time was 10.7~ks.  Data were taken with the
Proportional Counter Array (PCA) in {\tt GoodXenonwithPropane} mode,
which records photon arrival times with 1-$\mu$s resolution, and bins
photon energies into 256 channels.  In subsequent analysis, photon
arrival times at each epoch were adjusted to the solar system
barycentre.  The resulting time series were analysed in a variety of
ways.  Figure~1 shows the light curve binned with 125~ms time
resolution, along with time series of several properties of the pulsed
and persistent emission (see below).  The decreasing burst rate and
flux throughout our observation clearly indicates that we observed only
the end of an event that commenced prior to the start of our
observations.  Lightcurves in the three operational PCUs look similar.
Only the largest burst showed any excess in the PCA Standard 1
``Remaining Counts,'' however, the flux correction due to deadtime is
minimal ($\sim$10\%).

To identify all bursts, the time series made using photons in the
2--20~keV range were searched for significant excursions from the mean
count rate by comparing each time bin value with a windowed 7-s running
mean.  Bursts were identified assuming Poissonian statistics, and by
combining probabilities from the separate PCUs.  A detailed description
of the burst search algorithm and subsequent analysis will be presented
elsewhere.  Here we summarize the main results.  The bursts range in
duration from 2~ms to 3~s and show a variety of structures.  A few
($\sim$10) have clear fast-rise, exponential-tail morphology, however
the rest appear approximately symmetric.  Their spectra are all well
modeled by simple power laws, with indices in the range $\sim$0.25 to
$\sim$2.4.  Burst fluences in the 2--10 keV band are in the range
$3\times 10^{-11}$ to $5\times 10^{-9}$~erg~cm$^{-2}$, corresponding to
energies $3 \times 10^{34}$ to $5 \times 10^{36}$~erg, assuming
isotropic emission and a distance of 3~kpc to the source
\citep{kuy02}.  The sum total of all burst fluences is $3.2 \times
10^{-8}$~erg~cm$^{-2}$, corresponding to energy $3.4 \times
10^{37}$~erg (2--10 keV).  Peak burst fluxes in the 2--10~keV band are
in the range $1 \times 10^{-9}$ to $4 \times 
10^{-7}$~erg~s$^{-1}$~cm$^{-2}$, corresponding to peak luminosities of
$1 \times 10^{36}$ to $4 \times 10^{38}$~erg~s$^{-1}$.

Follow-up {\it RXTE} observations on June 20 revealed no further
bursts, nor have any of the 15 observations, each of duration 7--8~ks,
obtained every $\sim$10 days since.  Neither target-of-opportunity
observations obtained with the {\it XMM-Newton} satellite on June 21,
nor {\it XMM-Newton} observations scheduled fortuitously 7 days prior
to the burst, revealed any additional bursts (work in preparation).
The {\it RXTE} All Sky Monitor observed the field on June 18 at UTs
03:50 and 14:43 for $\sim$90~s per observation but detected no enhanced
flux, with 99\% confidence upper limits of $1\times
10^{-9}$~erg~s$^{-1}$~cm$^{-2}$ (2--10~keV).

In spite of the large (1$^{\circ}$) field of view of the PCA, we are
certain that the AXP is the origin of the bursts, as many properties of
the pulsed emission were simultaneously observed to change dramatically
(Fig. 1).  The persistent flux evolution was determined as follows.  A
spectral analysis was done using the {\tt XSPEC} software package
v11.2.0 \footnote{http://xspec.gsfc.nasa.gov}, in which the preburst
PCA data for 1E~2259+586 were modeled using the best available
background models made with the {\tt FTOOL pcabackest}
(http://heasarc.gsfc.nasa.gov/docs/xte/recipes/p2.html), the pulsar
spectrum as determined using a {\it XMM-Newton} observation of the
pulsar made 1 week before the burst observation (work in
preparation), and an additional component to account for the remaining
emission in the PCA field-of-view.  The burst data were modeled by a
blackbody plus power-law component, while holding the pulsar equivalent
neutral hydrogen column density and the remaining emission model
fixed.  The resulting persistent fluxes are show in Figure~1 (second
panel from the top).  The pulsed flux evolution, also show in in Figure~1
(second panel from the top) was calculated by first folding
$\sim$200~s long data segments with the spin ephemeris (Table 1), then
summing the first six harmonics of the normalized Fourier powers of the
resulting pulse profiles.  The total 2--10~keV fluence over and above
the quiescent flux is $2 \times 10^{-6}$~erg~cm$^{-2}$, two orders of
magnitude above that in the bursts.  As is also seen in Figure~1, the
pulsar spectrum clearly hardened during the outburst, and relaxed back
toward the quiescent spectral paramters during the course of the observation.
The fitted blackbody radius remained approximately constant throughout.
Interestingly, the same cannot be said of the ratio of power-law to blackbody
flux; as is seen in the bottom panel of Figure~1, the latter continued
evolving away from the quiescent state during our observation.

A significant change in the pulse
morphology was observed at the burst epoch, as shown in Figure 2.
During the outburst, the amplitudes of the peaks relative to the pre-
and post-outburst  profiles are clearly reversed.  The relative phase
displayed above is that successfully used in our timing analysis.  The
profile change is similar in different energy bands.  This different
pulse profile persisted for at least 2 days following the outburst, and
gradually returned to its pre-outburst morphology after $\sim$6 days
(Gavriil et al. in preparation).

The star underwent a sudden spin up or ``glitch'' at the outburst
epoch.  For details regarding how the timing analysis was done, see
\citet{gk02}.  
For the
burst and immediate post-outburst data, the pulse profile changes
described above resulted in obvious phase jumps corresponding to
the two peaks being swapped in the cross-correlation.   
The glitch epoch was determined by requiring zero phase jump between pre-
and post-outburst ephemerides.  
The {\it RXTE} data obtained
during and after the burst are well characterized (rms residual 1.5\%
of the period for the full data set) by $\Delta \nu / \nu = (4.10 \pm
0.03) \times 10^{-6}$, similar to that observed in radio pulsar
glitches \citep{ls90}.  The best-fit glitch epoch is consistent at the
$<1\sigma$ level with having occurred during our observation.
Additionally, the spin-down rate can be modeled as having approximately doubled abruptly.  Precise spin parameters are given in Table 1.  
Residuals from the last $\sim$100~days of timing suggest that the
spin-down rate may have relaxed back to near its pre-burst value
by $\sim$60~days post-outburst, however additional observations
are required to confirm this.

Target-of-opportunity near-infrared observations were made using the
NIRI instrument at the 8-m {\it Gemini}-North telescope in Hawaii on
June 21 at UT 14:44 using a $K_s$ filter (0.15 $\mu$m wide centred on
2.15~$\mu$m).  The observation had total exposure 1530~s with
0.7$''$ seeing and light cirrus.  The data were reduced using the {\it
Gemini} {\tt IRAF} package and photometry performed using standard {\tt
IRAF} procedures.  The proposed infrared counterpart \citep{htvk01} of
1E~2259+586 had magnitude $20.36 \pm 0.15$, $1.33 \pm 0.22$~mag (factor
of $3.40^{+0.77}_{-0.62}$) brighter 3~days after the outburst than was
observed in a 2000 {\it Keck} telescope observation \citep{htvk01}.  
A second
{\it Gemini}/NIRI observation was obtained on June 28 at UT 14:51, with
900~s of exposure and 0.55$''$ seeing.  This time, the AXP counterpart
had faded to magnitude $21.14 \pm 0.21$, for a difference relative to
the 2000 {\it Keck} observation of $0.56 \pm 0.29$~mag (factor of
$1.67^{+0.52}_{-0.39}$ in brightness).
Photometric measurements on 7
reference objects in the field agreed with those obtained at {\it Keck} to
within 0.007 mag and 0.028 mag for the first and second nights,
respectively.

Target-of-opportunity radio observations were also made using the {\it
Very Large Array} in New Mexico on June 20, 2002.  The 1E~2259+586
field was observed for 2420~s in B array at a central observing
frequency of 1424.3 MHz.  After standard calibration, imaging and
cleaning using the {\tt MIRIAD} software package, an rms noise level of
15~$\mu$Jy/beam ($4.6^{\prime \prime} \times 3.9^{\prime \prime}$ beam)
was achieved.  No emission was detected.  We place a 3$\sigma$ upper
limit of $50$~$\mu$Jy on the radio flux for this epoch.

\section{Discussion}

The X-ray phenomenology we have observed in this major AXP outburst is
all reminiscent of that seen in SGR bursts.  The short bursts (Fig. 1)
are very similar to short SGR bursts. The long, thermally evolving tail
is similar to that seen in a handful of SGR bursts \citep{lwg+03}.  A
timing anomaly in SGR 1900+14 was seen at the time of the giant flare
in 1998 \citep{wkg+02}, as was a pulse profile change and enhanced
pulsed and persistent flux \citep{gkw+01}.  Thus, this AXP has shown
uniquely SGR-like bursting behavior.  1E~2259+586 showed the most
stable timing behavior of all AXPs in the 5.6~yr prior to this
event \citep{gk02}, while 1E~1048.1$-$5937, the only other AXP seen to
burst \citep{gkw02}, showed the least stable behavior, as well as the
hardest AXP spectrum \citep{kgc+01}.  Thus it seems any AXP can burst.

The properties of the outburst solve a number
of previously outstanding AXP problems.  A similar pulse profile change
was claimed previously in data for 1E~2259+586 from the {\it Ginga}
mission in 1989 \citep{ikh92}.  The archival {\it Ginga} data show no
evidence of bursts.  The {\it Ginga} observation probably took place
just after an outburst, consistent with the reported timing anomaly at
the same epoch \citep{ikh92}.  This suggests that such outbursts occur on
decade time scales \citep{hh99}.  In addition, previously reported large
X-ray flux variations in 1E~2259+586 and
1E~1048.1$-$5937 \citep{ikh92,bs96,opmi98} that were called into question
by the flux stability observed in the $\sim$5~yr prior to June
2002 \citep{gk02,tgsm02} are now more understandable as enhanced emission
due to bursting episodes.  Pulsations from the AXP candidate
AX~J1845$-$0258 have been detected only once, in spite of multiple
observations \citep{vgtg00}.  This
may have been following a similar outburst (though there is no evidence
for bursts in the archival {\it ASCA} data).

The 1E~2259+586 outburst likely resulted from a sudden event in the
stellar crust, such as a crustal fracture, which simultaneously
affected both the superfluid interior and the magnetosphere.  The large
spin-up can be explained by the coupling of the faster-rotating
superfluid inside the star with the crust, following the unpinning of
angular momentum vortices from crustal nuclei.  The fractional
frequency increase is similar to that observed in many radio
pulsars \citep{ls90} but is smaller than could have been detected in SGR
timing data.  However, the factor of two increase in the spin-down rate
is unprecedented for radio pulsars, though possibly not for
SGRs \citep{wkg+02}.  According to glitch theory \citep{accp93}, when the
glitch occurs, a portion of the superfluid decouples, decreasing the
effective moment of inertia of the star.  For fixed external torque, an
increase in spin-down rate results.  For radio pulsars, the decoupled
portion amounts to $\sim$1\% of the stellar moment of inertia,
corresponding to the observed $\sim$1\% increases in spin-down
rates \citep{accp93}.  For 1E~2259+586, however, most of the stellar
moment of inertia would have had to decouple.  This could imply a
decoupling of core, as opposed to crustal, superfluid.

Alternatively, the external torque could have changed, due to a
restructuring of the magnetosphere.  Indeed the enhanced X-ray
luminosity is too large to be explained as energy dissipated by
vortex unpinning \citep{td96a} or crustal elastic energy \citep{rud91c}.  A
decaying magnetar-strength magnetic field can cause severe stress on
the crust.  A large-scale fracture could trigger vortex unpinning, and,
simultaneously, shift magnetic field footpoints, resulting in a
magnetospheric reconfiguration \citep{td95}.  The pulse profile variation
is unlikely to be a result of the change in magnetospheric
structure \citep{tlk02}, since the torque change is much longer-lived.
Rather, the profile change probably occurred at the surface; the
effective blackbody radius of $\sim$1~km as determined from the
spectral fits supports a localized enhancement.

Notably, there is no evidence for an accompanying giant soft gamma-ray
flare, as might be expected from a sudden restructuring of the surface
magnetic field of a magnetar \citep{wkv+99}.  From the {\it
Interplanetary Network} spacecraft, an upper limit on the fluence of a
soft gamma-ray flare from 1E~2259+586, near the time of the X-ray
outburst, is $5 \times 10^{-7}$~erg~cm$^{-2}$ (25-150 keV) on time
scales 0.25-0.5 s (K. Hurley 2002, personal communication).  This
corresponds to an energy $5 \times 10^{38}$~erg, six orders of
magnitude below that released in the giant SGR flares.  This is
consistent with the absence of any radio emission post-outburst from
1E~2259+586, as well as with the absence of bright soft gamma-ray
flares from this source in the past.  Thompson \& Duncan (1996) showed
that the absence of a large flare in the event of a glitch requires the
neutron star crust to have deformed plastically.  This demands a magnetic
field roughly two orders of magnitude greater than that implied by the
spin-down of 1E~2259+586.  This could be explained by higher order
multipole moments which are negligible at the light cylinder, where the
spin-down torque arises.

The characteristic age ($P/2\dot{P} \simeq 100-200$~kyr) of 1E~2259+586
is much larger than the inferred age ($\sim 10$~kyr) of the supernova
remnant CTB~109 in which it resides \citep{rp97}.   It is tempting to
explain this discrepancy as being due to the pulsar having episodes of
transient accelerated spin-down such as we observe post-outburst.
However at least in this instance, the increased spin-down rate could
be roughly compensated by the sudden spin-up.

The near-infrared enhancement post-outburst is intriguing.  Currently,
the magnetar model does not address the origin of such emission.  In
conventional rotation-powered pulsars, infrared emission is thought to
arise from a population of synchrotron radiating electron/positron
pairs in the outer magnetosphere (see \citet{ls90} and references
therein).  An enhancement is therefore consistent with a change in the
magnetospheric field structure suggested by the torque change.  Future
observations can test this by comparing infrared variations to the
torque evolution.

We thank M.~Lyutikov, R.~Manchester, and C.~Thompson for useful
discussions.  We thank F. Nagase and B. Paul for help with the archival
{\it Ginga} data.  This work was supported in part by NSERC, CIAR, and
NASA/LTSA and made use of data obtained through the HEASARC
Online Service, provided by the NASA/Goddard Space Flight Center.  We 
thank J. Swank and the {\it RXTE} scheduling team.
The {\it Gemini} Observatory is operated by the Association of
Universities for Research in Astronomy, Inc., under a cooperative
agreement with the NSF on behalf of the {\it Gemini} partnership: the
National Science Foundation (United States), the Particle Physics and
Astronomy Research Council (United Kingdom), the National Research
Council (Canada), CONICYT (Chile), the Australian Research Council
(Australia), CNPq (Brazil), and CONICET (Argentina).  We thank F.
Jansen for scheduling Target-of-Opportunity observations with {\it
XMM-Newton} and B. Clark for scheduling the Ad Hoc observation
with the VLA.  The National Radio Astronomy Observatory is a facility
of the National Science Foundation operated under cooperative agreement
by Associated Universities, Inc.


\newpage
\begin{deluxetable}{lc} 
\tablewidth{210pt}
\tablecaption{Spin Parameters for 1E 2259+586\tablenotemark{a}}
\tablehead{ \colhead{} & \colhead{}}
\startdata
No. TOAs & 112  \\
Range (MJD) & 50356--52575 \\
Epoch (MJD) & 52400.0000 \\
$\nu$ (Hz)  & 0.1432870351(3) \\
$\dot{\nu}$ ($\times 10^{-15}$ Hz s$^{-1}$) & $-9.811(8)$ \\
$\ddot{\nu}$ ($\times 10^{-24}$ Hz s$^{-2}$) & 1.28(9)  \\
Glitch Epoch (MJD) & 52443.9(2) \\
$\Delta \nu$ ($\times 10^{-7}$ Hz) & 5.88(4) \\
$\Delta \dot{\nu}$ ($\times 10^{-14}$ Hz s$^{-1}$) & $-$1.09(7) \\
rms Residual (ms) & 102 \\
\enddata
\tablenotetext{a}{Numbers in parentheses are {\tt TEMPO}-reported 1$\sigma$ uncertainties. } 
\end{deluxetable}

\clearpage
\begin{figure}
\columnwidth=0.65\columnwidth
\plotone{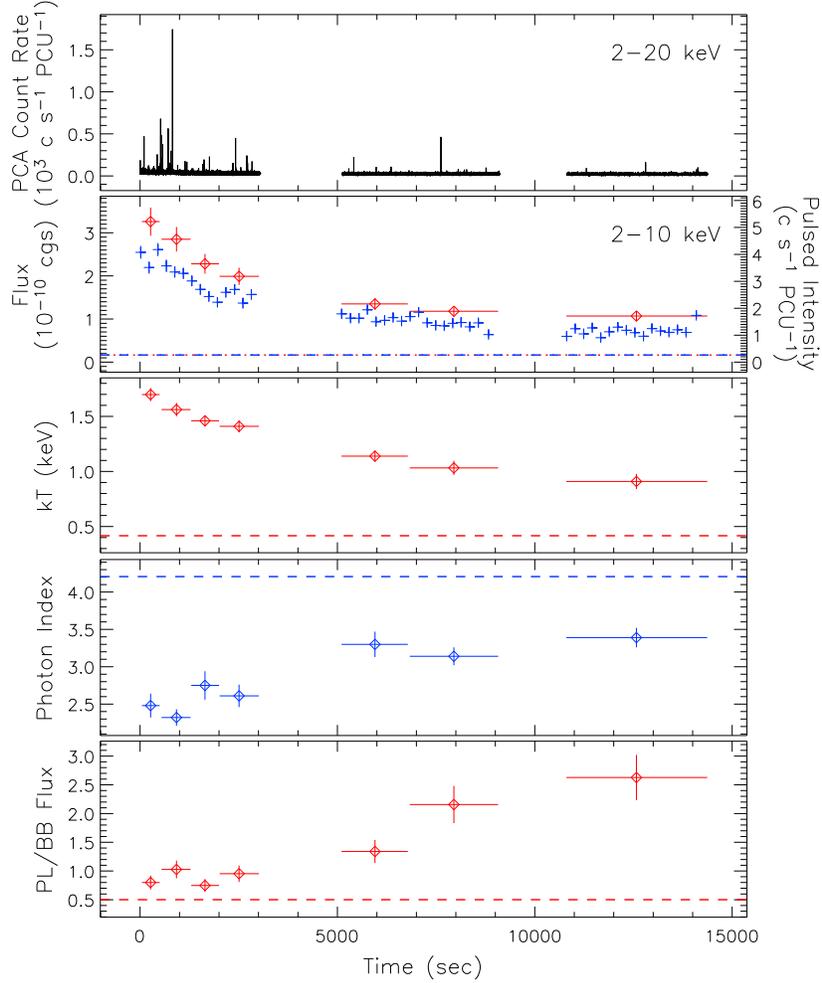}
\newpage
\figcaption{
Lightcurve and time evolution of persistent and pulsed emission during
the burst observation.  Top panel:  2--20~keV {\it RXTE/PCA} lightcurve
for 1E~2259+586 on June 18, 2002, at 125~ms resolution. 
The gaps are Earth occultations.  
2nd panel: Unabsorbed persistent (red diamonds) and
pulsed (blue crosses) fluxes in the 2--10 keV band.  The vertical scale
of each parameter has the same relative range to show the lower pulsed
fraction within this observation relative to the pre-burst value.  
The horizontal dashed (dotted)
lines denote the quiescent (pre-burst) levels of each parameter.  
3rd panel: Blackbody temperature of the persistent and pulsed emission
spectrum assuming a two-component model consisting of the blackbody and
a power law.  The same spectral fits show that the blackbody radius
remained at $\sim$1~km throughout.  4th panel: Power-law photon index
of the persistent and pulsed emission spectrum for same model as in the
3rd panel. 5th panel:  Ratio of the unabsorbed 2--10~keV power-law flux
and the bolometric blackbody flux.  
}
\label{fig:lc_pf}
\end{figure}

\clearpage
\begin{figure}
\plotone{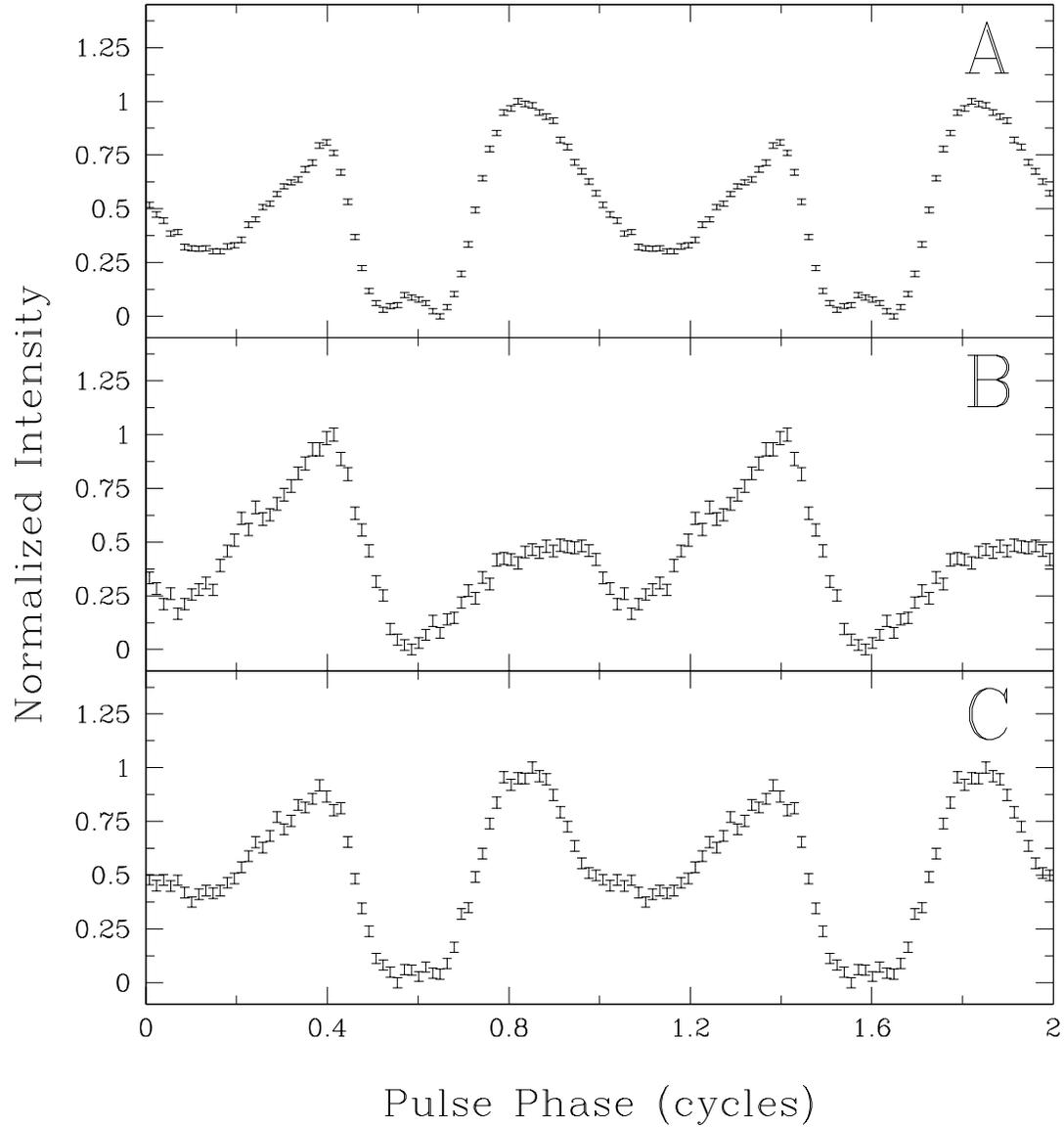}
\figcaption{
Average pulse profiles of 1E~2259.1+586 in the 2.5-9.0~keV band.  Two
cycles are plotted for clarity.  (A) Average profile before the
outburst (total exposure time: 764~ks).  (B) Average profile during the
outburst, with  bursts omitted (total exposure time: 11~ks).  (C)
Average profile beginning 12 days after the outburst (total exposure
time: 108~ks).  }
\label{fig:pulse_morph}
\end{figure}


\end{document}